\documentclass[aps,psfig,epsfig,preprint] {revtex4}
\usepackage{epsfig}
\usepackage{color}
\usepackage{bm}
\usepackage{subfigure}

\begin{document}
\draft
\title{
{\normalsize \hskip4.2in USTC-ICTS-06-15} \\{\bf A Candidate for
$1^{--}$ Strangeonium Hybrid}}

\author{Gui-Jun Ding\footnote{e-mail address: dinggj@mail.ustc.edu.cn},
Mu-Lin Yan\footnote{e-mail address: mlyan@ustc.edu.cn}}

\affiliation{\centerline{Interdisciplinary Center for Theoretical
Study and Department of Modern Physics,} \centerline{University of Science and Technology of
China,Hefei, Anhui 230026, China} }

\begin{abstract}
We propose that the recently observed structure at 2175MeV by the
Babar Collaboration is a $1^{--}$ strangeonium hybrid, and we
investigate this interpretation from both the flux tube model and
the constituent gluon model. The decay patterns and decay width in
the flux tube model and the constituent gluon model (for the "gluon
excited" hybrid) are very similar. The tetraquark hypothesis is not
favored by the available experimental data. The crucial test of our
scenario is suggested, furthermore, the promising channels which can
discriminate the hybrid interpretation from the tetraquark are also
suggested.

\end{abstract}
\maketitle
\section{introduction}

Very recently the Babar Collaboration observed a structure at
${\mathbf{2175}}$ MeV in the $\phi f_0(980)$ recoil mass of the
process $e^{+}e^{-}\rightarrow\gamma_{ISR}\phi
f_0(980)$\cite{babar}. Its mass is $m=2.175\pm0.010\pm0.015
\rm{GeV/c^2}$ and width is narrow $\Gamma=58\pm16\pm20 \rm{MeV}$. It
is claimed as a isospin singlet, and its spin-parity is determined
to be $J^{PC}=1^{--}$. Henceforth, this structure is denoted by
$Y$(2175). There are no known meson resonances with $I=0$ near this
mass, therefore it may not be a standard meson but rather a exotic.

Since both this structure and $Y$(4260)\cite{y4260} are observed in
$e^{+}e^{-}$ annihilation through initial state radiation, it may
have an analogical structure of $Y$(4260) with $c\bar{c}$ replaced
by $s\bar{s}$. $Y$(4260) has been interpreted as $c\bar{c}g$
hybrid\cite{y1,pene1}, a $cs\bar{c}\bar{s}$ tetraquark
state\cite{tetraquark} or others\cite{other}, therefore $Y$(2175)
maybe a $s\bar{s}g$ hybrid or a  $ss\bar{s}\bar{s}$ tetraquark
state. In Ref.\cite{wangzg}, the mass of $Y$(2175) which is taken as
a four quark state has been calculated from QCD sum rule. In the
tetraquark picture, although $Y$(2175) can not decay into
$\eta\eta$, $\eta\eta'$, $\eta'\eta'$ and $\phi\phi$ due to the
$C-$parity and $G-$parity, no symmetry forbids the decay
$Y(2175)\rightarrow\eta(\eta')\phi$, which can occur through the
so-called "fall-apart" mechanism\cite{fallapart}. With enough phase
space, the decay width is generally large. Moreover,
$\eta(\eta')\phi$ should be one of the dominant decay modes, which
is also against the experimental observation. In addition we would
like to mention that from the view of quark correlation the
tetraquark hypothesis prefer to have the diquark-antidiquark
picture, i,e. \{ss\}\{$\bar{s} \bar{s}$\}, where the diquark(or
antidiquark) is the so called "bad" diquark ("good" and "bad"
diquarks in Jaffe's terminology\cite{jaffe}). In short, four quark
interpretation of $Y$(2175) faces great challenge. However, more
experimental investigation is needed to completely exclude the
possibility of $Y$(2175) being a tetraquark, although the available
data already disfavors the tetraquark hypothesis.

Hybrid state is one of the most promising new species of hadrons,
extensive investigations in searching of the hybrid states have been
pursued, especially in the light hadrons. Although there is
amounting evidence for hybrid consisting of light quarks, they still
await confirmation, {\it e.g.,} both $\rho_1(1450)$ and
$\pi_1(1600)$ are good hybrid candidates. Hybrid states have been
studied in various approaches such as the flux tube
model\cite{isgur,close1,close2,page,new}, the constituent gluon
model\cite{dm,oliver,tanimoto}, the QCD sum rule\cite{hqsr} and the
lattice QCD\cite{lattice}. In the following, we will argue the
mysterious $Y$(2175) could be a $1^{--}$ strangeonium hybrid
($s\bar{{s}}g$) from the flux tube model and the constituent gluon
model.

In the flux tube model, a hybrid state is a quark-antiquark pair
moving on an adiabatic surface generated by an excited gluonic
flux-tube, and a hybrid meson would decay because of
phenomenological pair production described by $^3P_0$ model coupled
with a flux tube overlap. Isgur {\it et al.,} estimated that the
lowest lying strangeonium hybrid($s\bar{{s}}g$) could have a mass
about 1.9GeV\cite{isgur}. Close {\it et al.,} improved the previous
treatments using a Hamiltonian Monte Carlo algorithm, they predicted
that the mass of the lightest $s\bar{{s}}g$ hybrid is
2.1-2.2GeV\cite{close1}, which is consistent with experimental value
of $Y$(2175)($m=2.175\pm0.010\pm0.015 \rm{GeV}$). In the constituent
gluon model, a hybrid state is a quark-antiquark state with an
additional constituent gluon, such a meson would decay though gluon
dissociation into a $q\bar{q}$ pair\cite{dm,oliver,tanimoto}. The
constituent gluon is expected to add $0.7\sim1$GeV to the
corresponding quarkonia and the excitation energy is about 0.4 GeV
for the excited hybrid state, naively the mass of the
$\rm{s\bar{s}g}$ would be about 2.12-2.42GeV, which is also
consisitent experimental data of $Y$(2175). So the predictions for
mass of lightest strangeonium hybrid in both models supports that
$Y$(2175) could be a $s\bar{s}g$ hybrid state.

Since the quark pair are created from gluon in the constituent gluon
model, this model belong to the $^3S_1$ decay model\cite{3s1}.
Although $^3P_0$ model is more successful than the $^3S_1$ decay
model in conventional meson and baryon decays\cite{swanson}, we
don't know whether the same is also true in the hybrid decay.
Moreover the constituent gluon model is widely used to discuss the
decay of the hybrid meson, it is regarded as a useful theoretical
tool to study hybrid hadron, so we would like to investigate the
hybrid hypothesis from both the flux tube model and the constituent
gluon model. If the results in two models are consistent within the
uncertainties of both models, it will be a strong support to our
picture.

Extensive and thorough analyse has been done for the $1^{--}$
system, the existence for $1^{--}$ hybrids in both isovector and
isoscalar channels is required in $e^{+}e^{-}$ annihilation and
$\tau$ decay\cite{ayu,pdg}. We claim that $Y$(2175) is the isoscalar
$s\bar{{s}}g$ hybrid state of this $1^{--}$ multiplet, and other
candidates in this multiplet, such as $q\bar{s}g$,
$s\bar{q}g$($q=u,d$) are expected to be observed in future. In this
work, we will investigate the $1^{--}$ strangeonium hybrid from both
the flux tube model and constituent gluon model in details. The
outline of the paper is as follows. In section II, we study the
$1^{--}$ strangeonium hybrid from the flux tube model, and in
section III, this hybrid state is investigated in the constituent
gluon model. A brief summary and discussions are given in Section
IV.

\section{$1^{--}$ strangeonium hybrid from the flux tube model}

The flux tube model is extracted from the strong coupling limit of
the QCD lattice Hamiltonian, and decay occurs when the flux-tube
breaks at any point along its length, producing a $q\bar{q}$ pair in
a relative $J^{PC}=0^{++}$ state. Since we have to consider the
dynamics of the flux tube, the flux tube overlap has to be included
in addition to the color, flavor, spin and spatial overlap. Hybrid
decay has been studied carefully in the flux tube model and its
extended version for both the light flavor and heavy
flavors\cite{isgur,close2,page}. In the following, the strangeonium
hybrid decay will be considered using simple harmonic oscillation
approximation, we will follow the formalism of Close {\it et
al.,}\cite{close2}. This is typical of decay calculation and it has
been demonstrated that using Coulomb+linear wavefunctions from the
relativized quark model of Godfrey and Isgur does not change the
results significantly\cite{norma,godfrey}. In the narrow width
approximation, the partial decay width for the process $A\rightarrow
BC$ is,
\begin{equation}
\label{1}\Gamma_{LJ}(A\rightarrow
BC)=\frac{p_B}{(2J_A+1)\pi}\frac{\widetilde{M}_B\widetilde{M}_C}{\widetilde{M}_A}|M_{LJ}(A\rightarrow
BC)|^2
\end{equation}
where the phase space normalization of Kokoski and Isgur is
employed\cite{norma,godfrey}, $\widetilde{M}_A,
\widetilde{M}_B,\widetilde{M}_C$ are respectively the 'mock meson'
masses of $A,\;B,\;C$. $M_{LJ}(A\rightarrow BC)$ is the partial wave
amplitude for the decay process, $L$ is the relative angular
momentum between $B$ and $C$, $J$ is the total angular momentum of $B$ and $C$,
and $p_B$ is $B$'s momentum in the rest frame of the particle $A$. The hybrid wavefunction is taken as:
\begin{equation}
\label{2}\psi_A(\mathbf{r})=\sqrt{\frac{3\beta_A^{3+2\delta}}{2\pi\Gamma(3/2+\delta)}}\;r^{\delta}\;{\cal
D}^{1}_{M^{A}_L\Lambda}(\phi,\theta,-\phi)\exp(-\beta^2_A\mathbf{r}^2/2)
\end{equation}
Here $\delta=0.62$, $M^{A}_L$ is the projection of the orbital
angular momentum along $\mathbf{z}$ axis, and $\Lambda$ is the flux
tube angular momentum along the quark-antiquark axis in the hybrid.
In our case $M^{A}_L=0,\pm1$ and $\Lambda=\pm1$. The S.H.O.
wavefunctions for the angular momentum $L=0$ and $L=1$ ordinary
mesons respectively are:
\begin{equation}
\label{3}\psi_S(\mathbf{r})=\frac{\beta^{3/2}}{\pi^{3/4}}\exp(-\beta^2\mathbf{r}^2/2),~~~~~\psi_P(\mathbf{r})=\frac{2\sqrt{2}\beta^{5/2}}{\sqrt{3}\pi^{1/4}}\;rY_{1M}(\hat{\mathbf{r}})\exp(-\beta^2\mathbf{r}^2/2)
\end{equation}
where $\beta$ is the harmonic oscillation parameter, which can be
different for various mesons. The amplitude for the hybrid state
decaying into $S+P$-wave meson pair can be calculated analytically
using the S.H.O. wavefunction. The $1^{--}$ $s\bar{s}g$ hybrid can
decay into $K^*_2(1430)K$ with the relative angular momentum between
the final states being $\mathbf{2}$, and it can also decay into
$K_1(1270)K$ and $K_1(1400)K$ in $S$-wave and $D-$wave. Under the
first order approximation $\beta_B=\beta_C$, the corresponding decay
amplitudes are in the followings\cite{close2}:
\begin{eqnarray}
\nonumber &&M_{S1}((s\bar{s}g)\rightarrow K_1(1270)K)=\varrho~F((s\bar{s}g)\rightarrow K_1(1270)K)\frac{1}{3}(-3h_0+g_1-4h_2)\\
\nonumber &&M_{S1}((s\bar{s}g)\rightarrow K_1(1400)K)=\varrho~F((s\bar{s}g)\rightarrow K_1(1400)K)\frac{\sqrt{2}}{3}(-3h_0+g_1-4h_2)\\
\nonumber &&M_{D1}((s\bar{s}g)\rightarrow K_1(1270)K)=\varrho~F((s\bar{s}g)\rightarrow K_1(1270)K)\frac{\sqrt{2}}{6}(g_1+5h_2)\\
\nonumber &&M_{D1}((s\bar{s}g)\rightarrow K_1(1400)K)=\varrho~F((s\bar{s}g)\rightarrow K_1(1400)K)\frac{1}{3}(g_1+5h_2)\\
\label{4} &&M_{D2}((s\bar{s}g)\rightarrow
K^*_2(1430)K)=\varrho~F((s\bar{s}g)\rightarrow
K^*_2(1430)K)\frac{1}{\sqrt{2}}(g_1+5h_2)
\end{eqnarray}
where
$\varrho=(\frac{a\tilde{c}}{9\sqrt{3}}\frac{1}{2}A^{0}_{00}\sqrt{\frac{fb}{\pi}})\frac{\kappa\sqrt{b}}{(1+fb/(2\tilde{\beta}^2))^2}
\sqrt{\frac{2\pi}{3\Gamma(3/2+\delta)}}\;\frac{\beta_A^{3/2+\delta}}{\tilde{\beta}}$
with
$\tilde{\beta}^2\equiv(\beta^2_B+\beta^2_C)/2=\beta^2_B=\beta^2_C$.
$F((s\bar{s}g)\rightarrow K_1(1270)K),\;F((s\bar{s}g)\rightarrow
K_1(1400)K),\;F((s\bar{s}g)\rightarrow K^*_2(1430)K)$ are the flavor
factors in the corresponding decay processes, in our case
$F((s\bar{s}g)\rightarrow K_1(1270)K)=F((s\bar{s}g)\rightarrow
K_1(1400)K)=F((s\bar{s}g)\rightarrow K^*_2(1430)K)=2$. The
analytical expressions for $h_0,\;g_1,\;h_2$ are listed in the
Appendix A. To derive this result, $K_1(1270)$ and $K_1(1400)$ are
taken to be linear combinations of $^1P_1$ and $^3P_1$
states\cite{page,norma},
\begin{eqnarray}
\nonumber&&|K_1(1270)\rangle=\sqrt{\frac{2}{3}}|^1P_1\rangle+\sqrt{\frac{1}{3}}|^3P_1\rangle\\
\label{5}&&|K_1(1400)\rangle=-\sqrt{\frac{1}{3}}|^1P_1\rangle+\sqrt{\frac{2}{3}}|^3P_1\rangle
\end{eqnarray}

It is well-known that the lowest lying hybrid does not decay to
identical mesons, but prefers to decay into $S+P$-wave meson
pair\cite{isgur,orsay}. This rule has been shown to be more general
than specific models\cite{page2}. Although decaying into $S+S$-wave
meson pair is not forbidden if the internal structures(${\it
i.e.,}\;\beta$) of the two $S-$wave meson differ, the width is
generally proportional to $(\beta^2_B-\beta^2_C)^2$, which is
usually small.

For the $1^{--}$ strangeonium hybrid decaying into two S-wave
mesons, the strangeonium hybrid can decay into $K^{*}(892)K$,
$\phi\eta$, $\phi\eta'$ in relative $P-$wave, the amplitude is of
the following form\cite{close2},
\begin{eqnarray}
\nonumber &&M_{P1}((s\bar{s}g)\rightarrow
BC)=i(\frac{a\tilde{c}}{9\sqrt{3}}\frac{1}{2}A^{0}_{00}\sqrt{\frac{fb}{\pi}})\frac{\kappa\sqrt{b}}{(1+fb/(2\tilde{\beta}^2))^2}
\;\Delta\sqrt{\frac{\pi}{3\Gamma(3/2+\delta)}}\;\frac{\beta_A^{3/2+\delta}(\beta_B\beta_C)^{3/2}}{\tilde{\beta}^5}\\
\label{6}&&\times3F((\rm{s\bar{s}g})\rightarrow BC){\it P}
\end{eqnarray}
where $\Delta\equiv\beta^2_B-\beta^2_C$,
$\tilde{\beta}^2\equiv(\beta^2_B+\beta^2_C)/2$, and
$F((s\bar{s}g)\rightarrow BC)$ is the flavor factor in the process
$(s\bar{s}g)\rightarrow BC$. $P$ is the overlap integral which can
be integrated out using the technique of Appendix A.
\begin{eqnarray}
\nonumber
P&=&\int^{\infty}_{0}dr\;r^{2+\delta}j_1(\frac{Mp_Br}{m+M})\exp(-(2\beta^2_A+\tilde{\beta}^2-(\frac{\Delta}{2\tilde{\beta}})^2)\frac{r^2}{4})\\
\label{7}&=&
\frac{Mp_B}{m+M}\;\frac{2^{3+\delta}\Gamma(2+\frac{\delta}{2})}{3[2\beta^2_A+\tilde{\beta}^2-(\frac{\Delta}{2\tilde{\beta}})^2]^{2+\delta/2}}\;{_1F_1}(2+\frac{\delta}{2},\frac{5}{2},-\frac{M^2p^2_B}{(m+M)^2[2\beta^2_A+\tilde{\beta}^2-(\frac{\Delta}{2\tilde{\beta}})^2]})
\end{eqnarray}
$\eta$ and $\eta'$ are taken to be "perfect mixing":
$\eta=\frac{1}{2}(u\bar{u}+d\bar{d})-\frac{1}{\sqrt{2}}s\bar{s}$,
$\eta'=\frac{1}{2}(u\bar{u}+d\bar{d})+\frac{1}{\sqrt{2}}s\bar{s}$.
The mixing angle is consistent with the one obtained from the
$\eta-\eta'$ mass matrix. The flavor factors respectively are
$F((s\bar{s}g)\rightarrow K^{*}K)=2$,
$F((s\bar{s}g)\rightarrow \phi\eta)={\it
F}((s\bar{s}g)\rightarrow \phi\eta')=\sqrt{2}$.

The hybrid state can decay into 2S+1S meson pairs in relative
$P-$wave, if it is not forbidden by the phase space, {\it e.g,} the
$1^{--}$ strangeonium hybrid can decay to $K^{*}(1410)K$, if the
problematic state $K^{*}(1410)$ is taken as a $2\;^3S_1$
state\cite{barnes1}. But the decay mode $K(1460)K$ is forbidden by
the "spin selection rule"\cite{page}, since $K(1460)$ is a
$2\;^1S_0$ strange quarkonium\cite{barnes1}. The amplitude for the
hybrid decaying to $2S+1S$ final state is as follows:
\begin{eqnarray}
\nonumber&&M_{P0}((s\bar{s}g)\rightarrow
BC)=-i(\frac{a\tilde{c}}{9\sqrt{3}}\frac{1}{2}A^{0}_{00}\sqrt{\frac{fb}{\pi}})\;\frac{\kappa\sqrt{b}}{(1+fb/(2\tilde{\beta}^2))^3}\sqrt{\frac{\pi}{3\Gamma(3/2+\delta)}}\frac{\beta_A^{3/2+\delta}(4\tilde{\beta}^4-\Delta^2)^{3/4}}{128\sqrt{3}\tilde{\beta}^{11}}\\
\label{add1}&&\times3 F((s\bar{s}g\rightarrow BC))\;P^{\prime}
\end{eqnarray}
The overlap integral $P^{\prime}$ is,
\begin{eqnarray}
\nonumber&&P^{\prime}=\int^{\infty}_{0}dr
\;r^{2+\delta}j_1(\frac{Mp_Br}{m+M})\{r^2\Delta(\Delta-2\tilde{\beta}^2)^2(bf+2\tilde{\beta}^2)(\Delta+2\tilde{\beta}^2)+8\tilde{\beta}^2[\;bf\Delta^2+2\Delta(-4bf+\\
\nonumber&&5\Delta)\;\tilde{\beta}^2-8bf\tilde{\beta}^4-16\tilde{\beta}^6]\}\\
\nonumber&&=\frac{Mp_B}{3(m+M)}\;\{\Delta(\Delta-2\tilde{\beta}^2)^2(bf+2\tilde{\beta}^2)(\Delta+2\tilde{\beta}^2)\;\frac{2^{5+\delta}\;\Gamma(3+\frac{\delta}{2})}{[2\beta_A^2+\tilde{\beta}^2-(\frac{\Delta}{2\tilde{\beta}})^2]^{3+\delta/2}}\;{_1F_1}(3+\frac{\delta}{2}\;,\frac{5}{2}\;,\\
\nonumber&&-\frac{M^2p^2_B}{(m+M)^2[2\beta^2_A+\tilde{\beta}^2-(\frac{\Delta}{2\tilde{\beta}})^2]})+\tilde{\beta}^2[\;bf\Delta^2+2\Delta(-4bf+5\Delta)\;\tilde{\beta}^2-8bf\tilde{\beta}^4-16\tilde{\beta}^6]\\
\label{add2}&&\times\frac{2^{6+\delta}\;\Gamma(2+\frac{\delta}{2})}{[2\beta_A^2+\tilde{\beta}^2-(\frac{\Delta}{2\tilde{\beta}})^2]^{2+\delta/2}}\;{_1F_1}(2+\frac{\delta}{2}\;,\frac{5}{2}\;,
-\frac{M^2p^2_B}{(m+M)^2[2\beta^2_A+\tilde{\beta}^2-(\frac{\Delta}{2\tilde{\beta}})^2]})\}
\end{eqnarray}

We take the string tension $b=0.18\rm{GeV}^2$, and the constituent
quark masses $m_u=m_d=0.33\rm{GeV}$, $m_s=0.55\rm{GeV}$, the masses
of the mesons will be taken from PDG\cite{pdg}. A detailed
discussion about other quantities such as $f$, $\kappa$,
$A^{0}_{00}$ may be found in the Appendix A of Ref.\cite{norma} and
Ref.\cite{paton}. The common factor
$\frac{a\tilde{c}}{9\sqrt{3}}\frac{1}{2}A^{0}_{00}\sqrt{\frac{fb}{\pi}}$
is taken to be 0.64 which gives a best fit to the decay of
conventional mesons in the flux tube model\cite{close2,norma}, and
the estimated values $f=1.1$, $\kappa=0.9$ is used in this work. The
oscillation parameter $\beta$ is chosen as that of
Ref.\cite{close2}, for the strangeonium hybrid $s\bar{s}g$ we choose
$\beta_A=0.30\rm{GeV}$. $\tilde{\beta}=0.40\rm{GeV}$ is used in the
case of hybrid decaying into $S+P-$wave mesons pair. For the
$S+S-$wave mesons final states, the effective $\beta $ in
Ref.\cite{norma} is used to determined the width, {e.g.}
$\beta_{K^{*}(1410)}=0.41\rm{GeV}$, $\beta_{K^{*}}=0.48\rm{GeV}$,
$\beta_{K}=0.71\rm{GeV}$, $\beta_{\eta}=\beta_{\eta'}=0.74\rm{GeV}$
and $\beta_{\phi}=0.51\rm{GeV}$. In the following table, we present
the dominant decay modes for the $1^{--}$ strangeness hybrid decay
in various partial waves.
\begin{table}[hptb]
\begin{center}
\caption{The decay modes of $1^{--}$ strangeonium hybrid $s\bar{s}g$ in the flux
tube model}
\begin{tabular}{|c|c|c|c|c|c|c|c|c|}\hline\hline
Decay Channels&  $K^{*}_2(1430)K$  & $K_1(1270)K$ & $K_1(1400)K$ &  $K^{*}(1410)K$   &$K^{*}(892)K$ & $\phi\eta$ & $\phi\eta'$& Total\\
\hline

$\Gamma(\rm{MeV})$ & D~~~15.0 & S~~~30.8& S~~~65.8& P~~~23.0  &P~~~3.7 & P~~~1.2 & P~~~0.4& 148.7\\
                   &          & D~~~4.5 & D~~~4.3 &   &      &         &        &
\\\hline\hline
\end{tabular}
\end{center}
\end{table}

From the above table, we can see the "S+P" selection rule is
satisfied for the strangeonium hybrid in a good manner, and the
hybrid decaying into $2S+1S$ final state is usually not suppressed.
The $1^{--}$ strangeonium hybrid mainly decays into $K_1(1400)K$,
$K_1(1270)K$, $K^{*}(1410)K$, $K^{*}_2(1430)K$. However, $Y$(2175)
mainly decays into $\phi\eta$ and $\phi\eta'$ in the
$s\bar{s}s\bar{s}$ tetraquark scenario, thus a search for the latter
channels, or the limit on its coupling, could be a significant
discriminator for the nature of $Y$(2175). Furthermore, the $1^{--}$
strangeonium hybrid can decay to $\phi\pi\pi$ by the cascade decay
mechanism
$(s\bar{s}g)\rightarrow(s\bar{s})(gg)\rightarrow\phi+\pi+\pi$\cite{cascade},
and $(s\bar{s}g)\rightarrow\phi f_0(980)$ may make an significant
contribution to the process. So the search for $\phi KK$ decay modes
are also merited.

\begin{figure}[hptb]
\centering
\begin{minipage}[t]{0.46\textwidth}
\centering
\includegraphics[width=8cm]{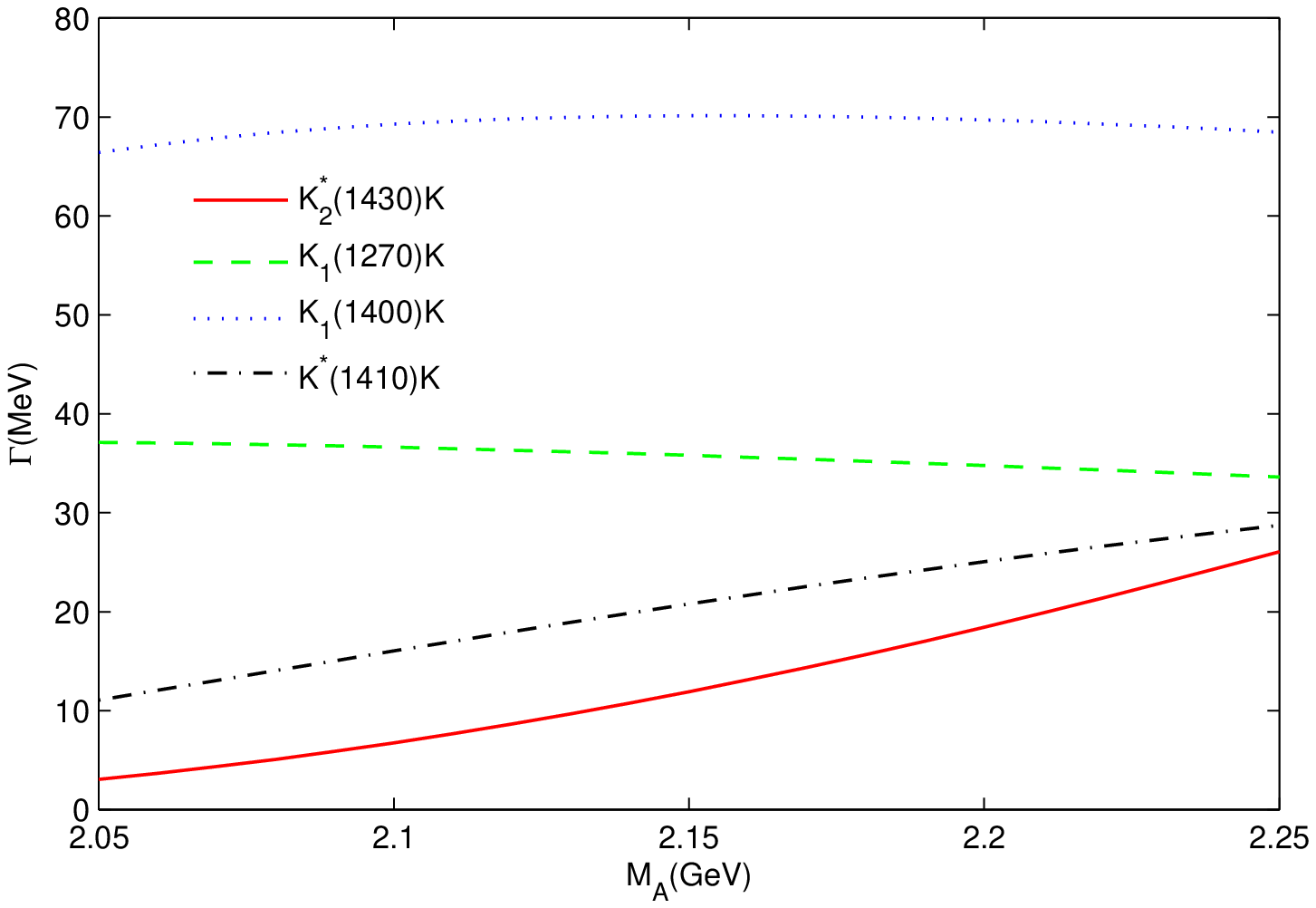}
\caption{The partial width for the $1^{--}$ strangeonium hybrid
decay as a function of the hybrid mass} \label{fig1}
\end{minipage}%
\hspace{0.04\textwidth}%
\begin{minipage}[t]{0.46\textwidth}
\centering
\includegraphics[width=8cm]{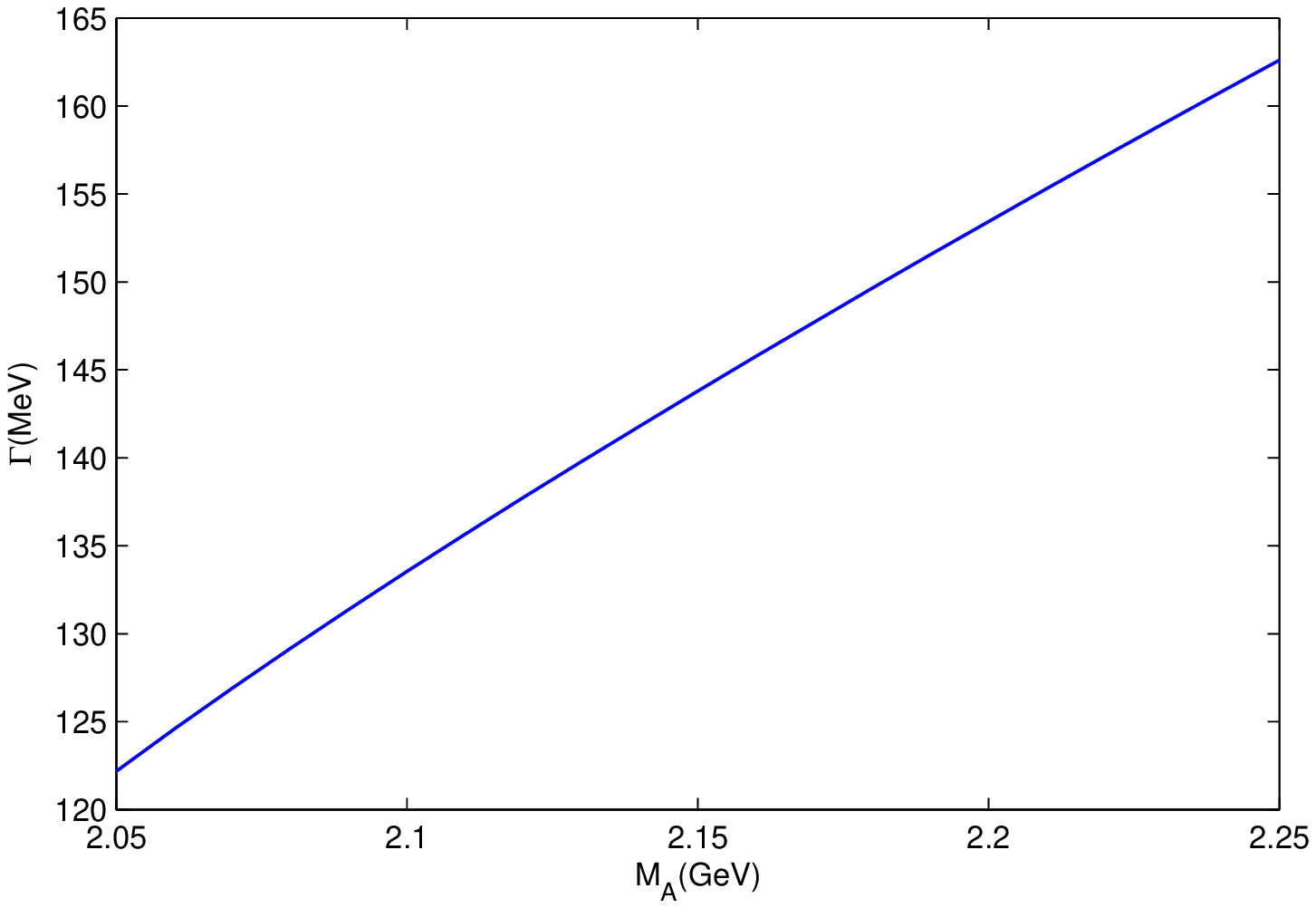}
\caption{The total width of the $1^{--}$ strangeonium hybrid as a
function of the hybrid mass} \label{fig2}
\end{minipage}
\end{figure}

\begin{figure}[hptb]
\centering
\begin{minipage}[t]{0.46\textwidth}
\centering
\includegraphics[width=8cm]{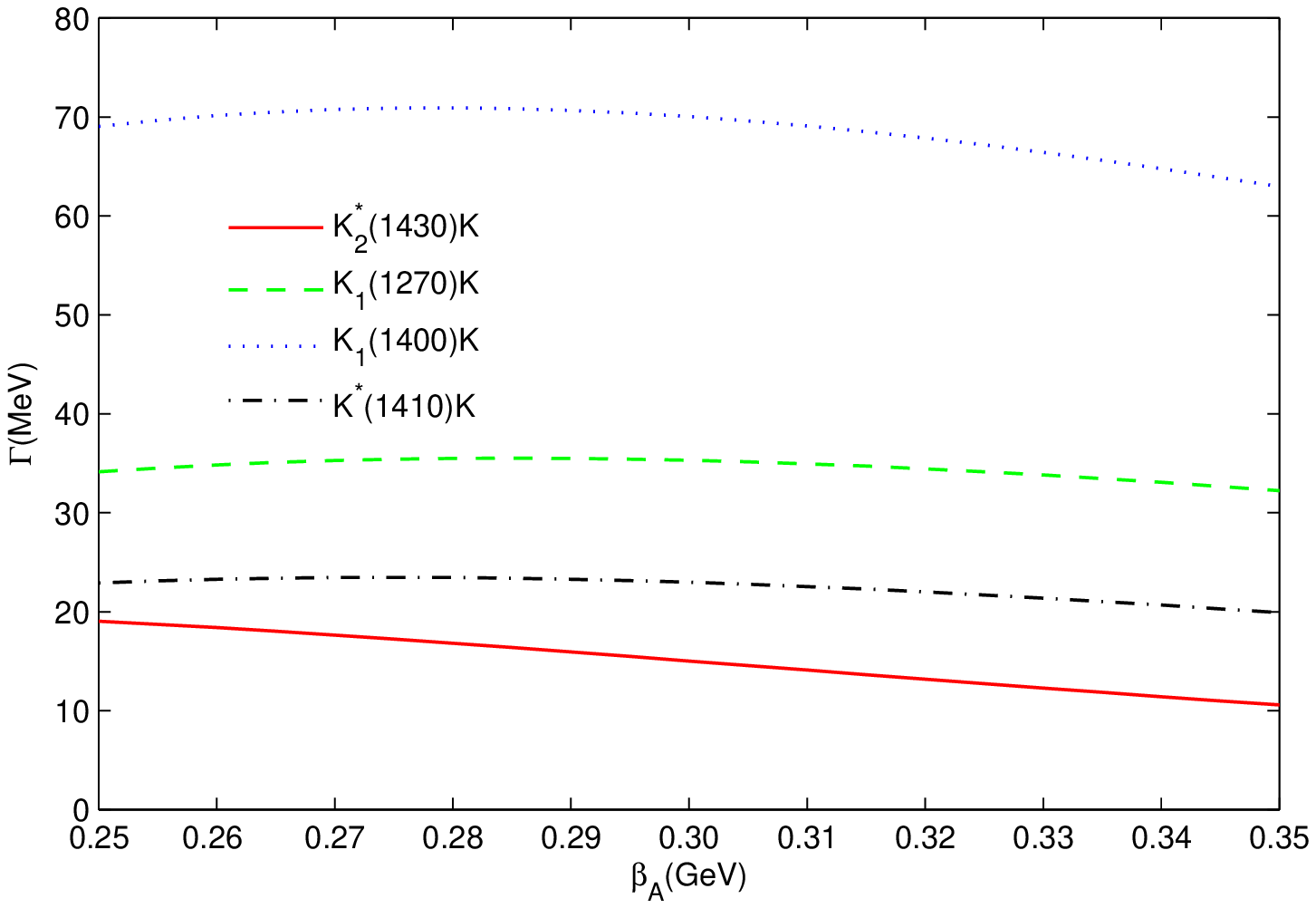}
\caption{The dependence of the partial width on $\beta_A$ for the
$1^{--}$ strangeonium hybrid decay} \label{fig3}
\end{minipage}%
\hspace{0.04\textwidth}%
\begin{minipage}[t]{0.46\textwidth}
\centering
\includegraphics[width=8cm]{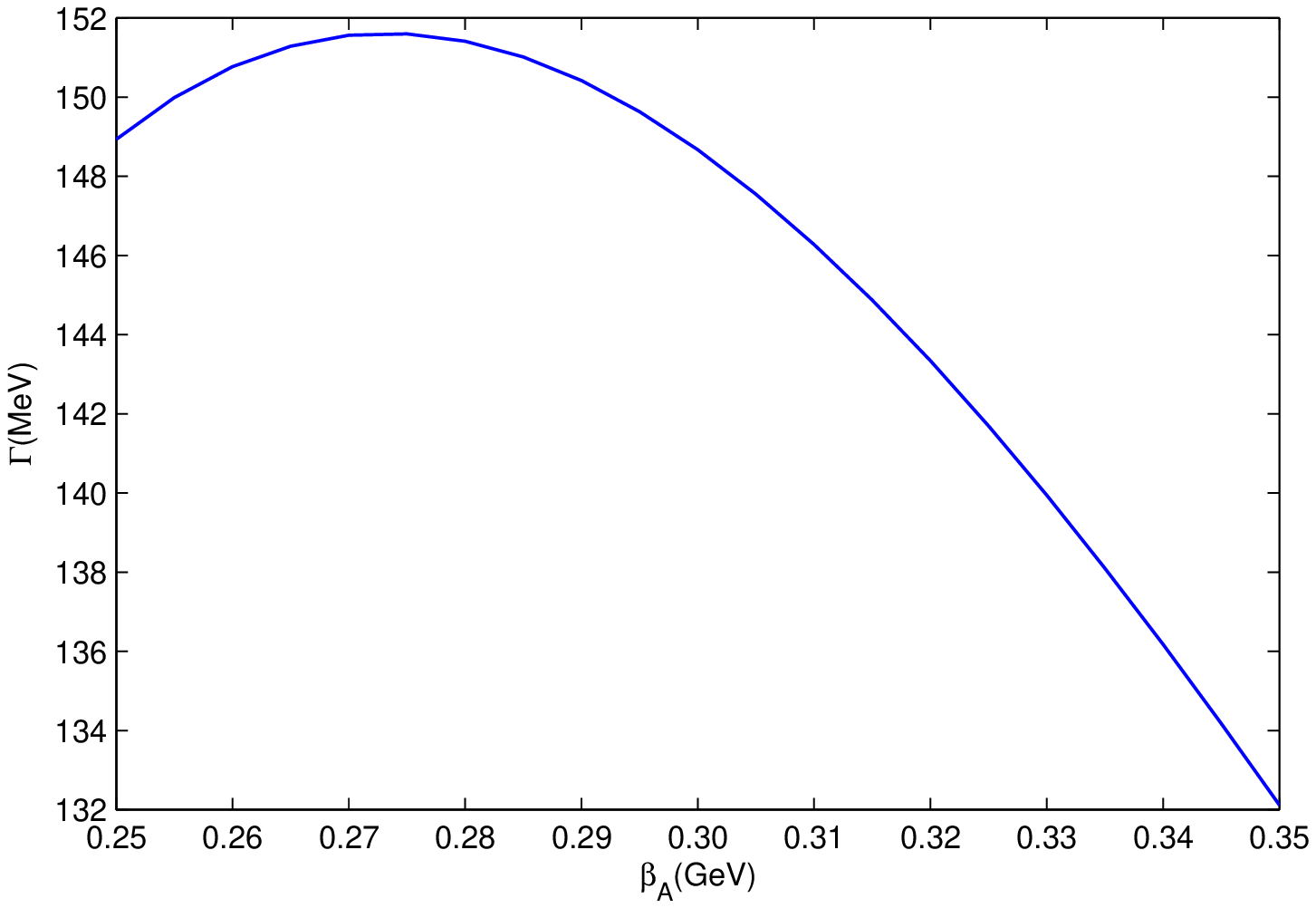}
\caption{The dependence of the total width of the $1^{--}$
strangeonium hybrid on $\beta_A$} \label{fig4}
\end{minipage}
\end{figure}

As a measure of the reliability of these predictions, the mass
dependence of the partial decay width and total width are
respectively shown in Fig.\ref{fig1} and Fig.\ref{fig2}. We also
show the harmonic parameter $\beta_A$ dependence of the partial
decay width and total width in Fig.\ref{fig3} and Fig.\ref{fig4}. We
don't show the partial width of the modes $K^*(892)K$, $\phi\eta$
and $\phi\eta'$, since they are small enough to be negligible. From
these figures, we can see the total decay width decreases with the
decrease of its mass. For a 2-2.2 GeV $1^{--}$ strangeonium hybrid,
its width is about 120-150MeV, which is consistent with the
experimental observation about $Y$(2175)($\Gamma=58\pm16\pm20
\rm{MeV}$) within the uncertainties of the flux tube model. On all
accounts, it is reasonable to identify $Y$(2175) as a $1^{--}$
strangeonium hybrid from the flux tube model.

\section{$1^{--}$ strangeonium hybrid from the constituent gluon model}

In order to provide further support to our proposal, we would like
to study the $1^{--}$ strangeonium hybrid from the constituent gluon
model. The constituent gluon model is a generalization of the quark
model with constituent gluon, the decay of hybrid mesons has been
studied widely in the constituent gluon
model\cite{oliver,tanimoto,orsay,congluon}. The reasonableness of
$Y$(4260) as a $1^{--}$ charmonium hybrid has been investigated from
this model\cite{pene1,iddir}. Here we will follow the notation of
Ref.\cite{oliver}. From the view of the constituent gluon model,
strangeonium hybrid is a bound state of $s\bar{s}$ and a gluon.
Defining ${L_g}$ as the relative orbital angular momentum between
the gluon and the $s\bar{s}$ center of mass, ${L_{s\bar{s}}}$ as the
relative angular momentum between the strange quark and the
anti-strange quark, ${S_{s\bar{s}}}$ as the total spin of
$s\bar{s}$. Denoting the total gluon angular momentum by
${\bf{J}}_g$, ${\bf{J}}_g\equiv {\bf{L}}_g+1$, and ${\bf{L}}\equiv
{\bf{L}}_{s\bar{s}}+{\bf{J}}_{g}$. The parity and charge conjugation
of the hybrid are given by:
\begin{equation}
\label{8}
P=(-1)^{L_{s\bar{s}}+L_{g}},~~~~C=(-1)^{L_{s\bar{s}}+S_{s\bar{s}}+1}
\end{equation}
The quantum number $J^{PC}=1^{--}$ implies that $L_{s\bar{s}}=S_{s\bar{s}}=1,\;L_g=0$ or
$L_{s\bar{s}}=S_{s\bar{}s}=0,\;L_g=1$, the former is usually is
referred as the "quark excited" hybrid, and the latter referred as the
"gluon excited" hybrid. To lowest order, the decay is described by the
matrix element of the QCD interaction between the initial hybrid
wave function and final two mesons wavefunctions, and the
interaction Hamiltonian is:
\begin{equation}
\label{9}H_I=g_s\int
d^3\mathbf{x}\;\bar{\psi}(\mathbf{x})\;\gamma_{\mu}\frac{\lambda^{a}}{2}\psi(\mathbf{x})\;A^{\mu}_a(\mathbf{x})
\end{equation}
The operator relevant to the decay can be expressed in terms of the
creation and annihilation operators:
\begin{equation}
\label{10}H_I=g_s\sum_{s,s^{\prime},\lambda,c,c^{\prime},c_{g}}\int\frac{d^3\mathbf{p}\;d^3\mathbf{p}^{\prime}d^3\mathbf{k}}{\sqrt{2\omega}(2\pi)^9}(2\pi)^3\delta^{3}
(\mathbf{p}-\mathbf{p^{\prime}}-\mathbf{k})\;\bar{u}_{\mathbf{p}sc}\gamma_{\mu}\frac{\lambda^{c_g}_{c,c^{\prime}}}{2}v_{-\mathbf{p}^{\prime}s^{\prime}c^{\prime}}b^{\dagger}_{\mathbf{p}sc}
d^{\dagger}_{-\mathbf{p}^{\prime}s^{\prime}c^{\prime}}a^{c_g}_{\mathbf{k}\lambda}\varepsilon^{\mu}(\mathbf{k},\lambda)
\end{equation}
here the flavor index of the quark has been omitted, $c_g=1,2,\cdot\cdot\cdot8$, and
\begin{eqnarray}
\nonumber&&\{b_{\mathbf{p}sc},b^{\dagger}_{\mathbf{p}^{\prime}s^{\prime}c^{\prime}}\}=\{d_{\mathbf{p}sc},d^{\dagger}_{\mathbf{p}^{\prime}s^{\prime}c^{\prime}}\}=(2\pi)^3\delta^3(\mathbf{p}-\mathbf{p}^{\prime})\delta_{ss^{\prime}}\delta_{cc^{\prime}}\\
\label{11}&&[a^{c_g}_{\mathbf{k}\lambda},a^{{c^{\prime}_g}^{\dagger}}_{\mathbf{k}^{\prime}\lambda^{\prime}}]=(2\pi)^{3}\delta^{3}(\mathbf{k}-\mathbf{k}^{\prime})\delta^{c_g,c^{\prime}_g}\delta_{\lambda\lambda^{\prime}}
\end{eqnarray}
under the non-relativistic approximation, the hybrid and the meson
states are described as follows:
%{M_{L_{g}},\lambda_g,M_{L_{s\bar{s}}},M_{S_{s\bar{s}}}}
\begin{eqnarray}
\nonumber&&|A(L_g,L_{s\bar{s}},S_{s\bar{s}},J_A,M_{J_A})\rangle=\sum\int\frac{d^3\mathbf{p}_1d^3\mathbf{p}_2d^3\mathbf{k}}{(2\pi)^9}
\langle L_{g},M_{L_{g}};1,\lambda_{g}|J_{g},M_{J_{g}}\rangle\langle
L_{s\bar{s}},M_{L_{s\bar{s}}};J_{g},\\
\nonumber&&M_{J_{g}}|L,m^{\prime}\rangle\langle
L,m^{\prime};S_{s\bar{s}},M_{S_{s\bar{s}}}|J_A,M_{J_{A}}\rangle(2\pi)^3\delta^{3}(\mathbf{p}_1+\mathbf{p}_2+\mathbf{k}-\mathbf{p}_A)\chi^{12}_{S_{s\bar{s},M_{S_{s\bar{s}}}}}\varphi^{12}_{A}\frac{\lambda^{c_{g}}_{c_1c_2}}{4}\\
\label{12}&&\psi_{L_{s\bar{s}}M_{L_{s\bar{s}}}}(\frac{\mathbf{p}_1-\mathbf{p}_2}{2})\;\psi_{L_{g}M_{L_g}}(\frac{2M\;\mathbf{k}-m_g
(\mathbf{p}_1+\mathbf{p}_2)}{2M+m_g})\;b^{\dagger}_{\mathbf{p}_1s_1c_1}d^{\dagger}_{\mathbf{p}_2s_2c_2}a^{{c_g}^{\dagger}}_{\mathbf{k}\lambda_g}|0\rangle
\end{eqnarray}
The subscript 1 and 2 refer to the strange quark and the
anti-strange quark within the hybrid meson. $\chi^{12}_{S_{s\bar{s},
M_{S_{s\bar{s}}}}},\;\varphi^{12}_{A}$ are respectively the spin
wavefunction and flavor wavefunction of $s\bar{s}$.
$\psi_{L_{s\bar{s}}M_{L_{s\bar{s}}}}(\frac{\mathbf{p}_1-\mathbf{p}_2}{2}),\;
\psi_{L_{g}M_{L_g}}(\frac{2M\mathbf{k}-m_g(\mathbf{p}_1+\mathbf{p}_2)}{2M+m_g})$
are respectively the spatial wavefunction of the $s\bar{s}$ and the
constituent gluon, which usually is taken as the simple harmonic
oscillator wavefunction.

The final $B$ meson's state is given by:
\begin{eqnarray}
\nonumber&&|B(L_B,S_B,J_B,M_{J_B})\rangle=\sum_{M_{L_B},M_{S_B}}\int\frac{d^3\mathbf{p}_1d^3\mathbf{p}_3}{(2\pi)^6}\langle L_{B},M_{L_{B}};S_{B},M_{S_{B}}|J_{B},M_{J_{B}}\rangle(2\pi)^3\delta^3(\mathbf{p}_1+\mathbf{p}_3\\
\label{13}&&-\mathbf{p}_{B})\;\chi^{13}_{S_BM_{S_B}}\varphi^{13}_B\;\omega^{13}_B\;\psi_{L_B
M_{L_{B}}}(\frac{m\mathbf{p}_1-M\mathbf{p}_3}{M+m})\;b^{\dagger}_{\mathbf{p}_1s_1}d^{\dagger}_{\mathbf{p}_3s_3}|0\rangle
\end{eqnarray}
Here $\omega^{13}_B$ is the color wavefunction of $B$ meson. Another
final state $C$ meson's wavefunction can be write out analogously in
terms of the quark, antiquark creation operators. It is
straightforward to get the matrix element $\langle
BC|H_I|A\rangle=g_s(2\pi)^3\delta^{3}(\mathbf{p}_A-\mathbf{p}_B-\mathbf{p}_C)\;M_{\ell,J}(A\rightarrow
BC)$, so the partial decay width is:
\begin{equation}
\label{18}\Gamma_{\ell J}(A\rightarrow
BC)=\frac{\alpha_s}{\pi}\frac{p_BE_BE_C}{M_A}|M_{\ell
J}(A\rightarrow BC)|^2
\end{equation}
where $M_{\ell J}(A\rightarrow BC)$ is the partial wave amplitude,
and $\Gamma_{\ell J}$ is the partial width of the corresponding
partial wave\cite{partialwave}. As usual, we would like to use the
S.H.O basis wavefunctions, thereby enabling analytic studies that
reveal the relationship among amplitudes. The partial wave amplitude
$M_{\ell J}$ and the spatial overlap for various final states are
presented in the Appendix B.

In our calculation, we take the following set of parameters:
$\beta_B=\beta_C=\beta_g=0.4\rm{GeV}$,
$\beta_{s\bar{s}}=0.3\rm{GeV}$, $m_s=0.55\rm{GeV}$,
$m_u=m_d=0.33\rm{GeV}$ and $m_g=0.8\rm{GeV}$, which are often used
in the constituent gluon model calculations\cite{orsay}. In the
Ref.\cite{orsay}, the authors used the oscillation parameters $R_B$,
$R_g$ and so on, the parameters $\beta_B$, $\beta_g$ etc in this
work are related to $R_B$, $R_g$ etc by the relations
$\beta_B=1/R_B$, $\beta_g=1/R_g$.

In the case of the "quark excited" hybrid, the allowed decay
channels and the partial decay width are shown in Table II. In this
table, $S$ or $P$ indicate that the relative angular momentum
between the two final states is $S$ or $P$ wave, and the symbols
0,1,2 denote the total angular momentum of two final states. From
this table, we can see the quark excited hybrid mainly decays into
$KK$, $K^{*}K$, $\phi\eta$, $\phi\eta^{\prime}$,
$K^{*}(892)K^{*}(892)$, $K_1(1270)K$, $K_1(1400)K$. The decay width
is very large, which is approximately 307.4MeV, 342.7MeV, 464.2MeV
respectively for the $L=0,1,2$ quark excited hybrid.

\begin{table}[hptb]
\begin{center}
\caption{Decay of the $1^{--}$ quark excited strangeonium hybrid
${s\bar{s}g}$ in the constituent gluon model, width is in
$\rm{MeV}\times$$\alpha_s$ for the channels. The QCD coupling
constant $\alpha_s$ is of order 1 in this nonperturbative region.}
\begin{tabular}{|c|c|c|c|}\hline\hline
                        &  L=0           &    L=1          &   L=2  \\\hline

$KK$                    & P~~~0~~~~14.3  & P~~~0~~~~43.0   &
P~~~0~~~~71.7\\\hline

$K^{*}K$                & P~~~1~~~~63.2 &  P~~~1~~~~47.4   &
P~~~1~~~~79.0\\\hline

$\phi\eta$              & P~~~1~~~~30.6 &  P~~~1~~~~22.9   &
P~~~1~~~~38.2\\\hline

$\phi\eta^{\prime}$     & P~~~1~~~~12.0 &  P~~~1~~~~9.0   &
P~~~1~~~~15.0\\\hline

                        & P~~~0~~~~4.5  & P~~~0~~~~13.4 &
P~~~0~~~~22.3\\
$K^{*}(892)K^{*}(892)$  & P~~~1~~~~0    & P~~~1~~~~0    &
P~~~1~~~~0\\
                        & P~~~2~~~~89.3 & P~~~2~~~~67.0 &
P~~~2~~~~4.5\\\hline

$K_1(1270)K$            & S~~~1~~~~28.9 & S~~~1~~~~54.2 &
S~~~1~~~~91.0\\

                        & D~~~1~~~~4.8  & D~~~1~~~~12.3 &
D~~~1~~~~~20.6\\\hline

$K_1(1400)K$            & S~~~1~~~~46.2 & S~~~1~~~~55.5 &
S~~~1~~~~92.0\\

                        & D~~~1~~~~1.8  & D~~~1~~~3.3 &
D~~~1~~~~~5.4\\\hline

$K^{*}_2(1430)K$        & D~~~2~~~~3.1  & D~~~2~~~~2.3 &
D~~~2~~~~3.9\\\hline

$K(1460)K$              & P~~~0~~~~2.6  & P~~~0~~~~7.8 &
P~~~0~~~~13.0\\\hline

$K^{*}(1410)K$          & P~~~1~~~~6.1  & P~~~1~~~~4.6 &
P~~~1~~~~7.6\\\hline

Tot                     &  307.4        & 342.7       &
464.2\\\hline\hline
\end{tabular}
\end{center}
\end{table}
\begin{table}[hptb]
\begin{center}

\caption{Decay of the $1^{--}$ gluon excited strangeonium hybrid
${s\bar{s}g}$ in the constituent gluon model, width is in
$\rm{MeV}\times$$\alpha_s$ for the channels. The QCD coupling
constant $\alpha_s$ is of order 1 in this nonperturbative region.}
\begin{tabular}{|c|c|c|c|}\hline\hline
$K_1(1270)K$  & $K_1(1400)K$  &  $K^{*}(1410)K$ &Total\\\hline\hline

36.2 & 73.2          &  5.8            &115.2
\\\hline\hline
\end{tabular}
\end{center}
\end{table}

\begin{figure}[hptb]
\centering
\begin{minipage}[t]{0.47\textwidth}
\centering
\includegraphics[width=8cm]{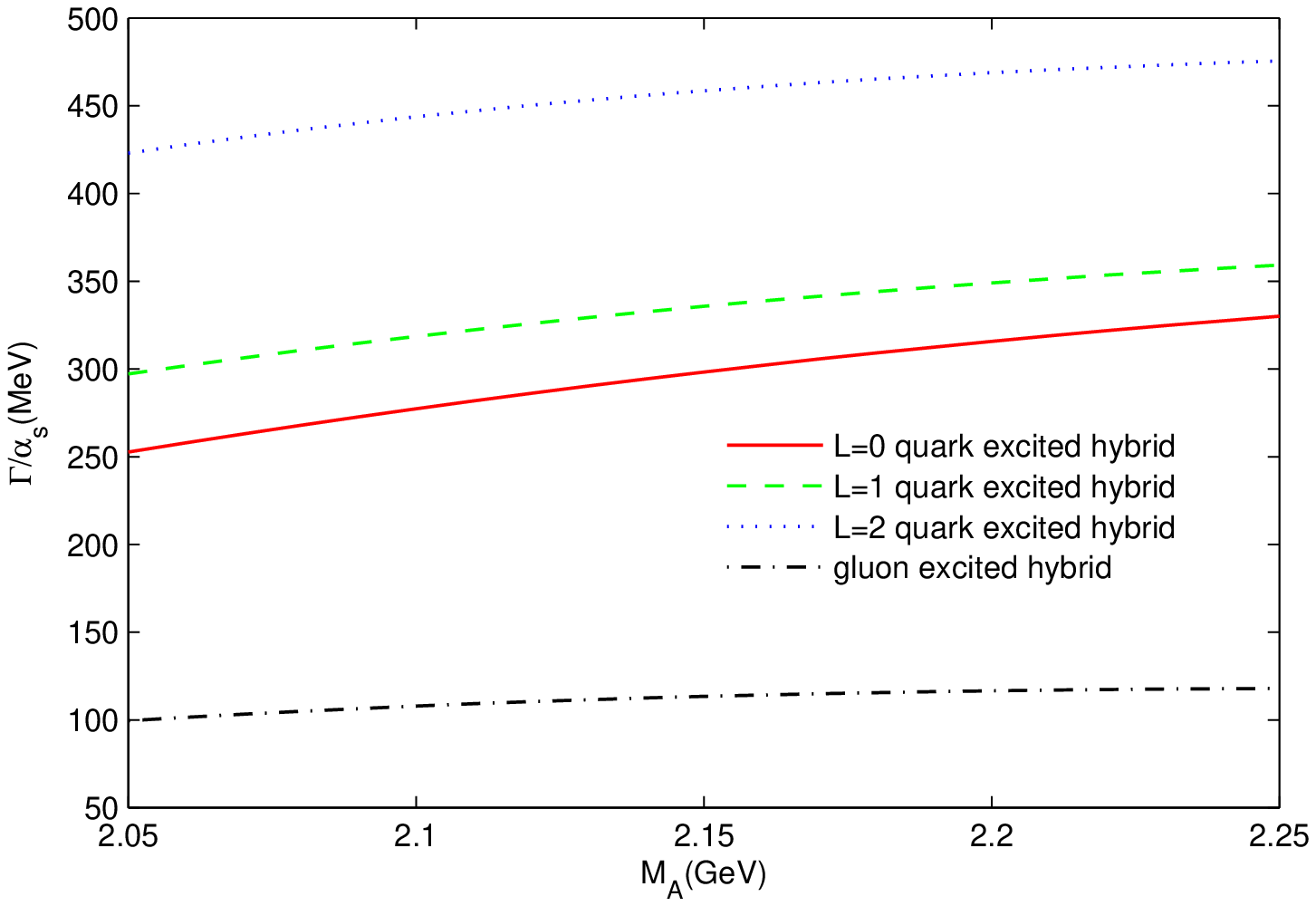}
\caption{The decay width for the $1^{--}$ strangeonium hybrid at
various hybrid mass.} \label{fig5}
\end{minipage}%
\hspace{0.03\textwidth}%
\begin{minipage}[t]{0.47\textwidth}
\centering
\includegraphics[width=8cm]{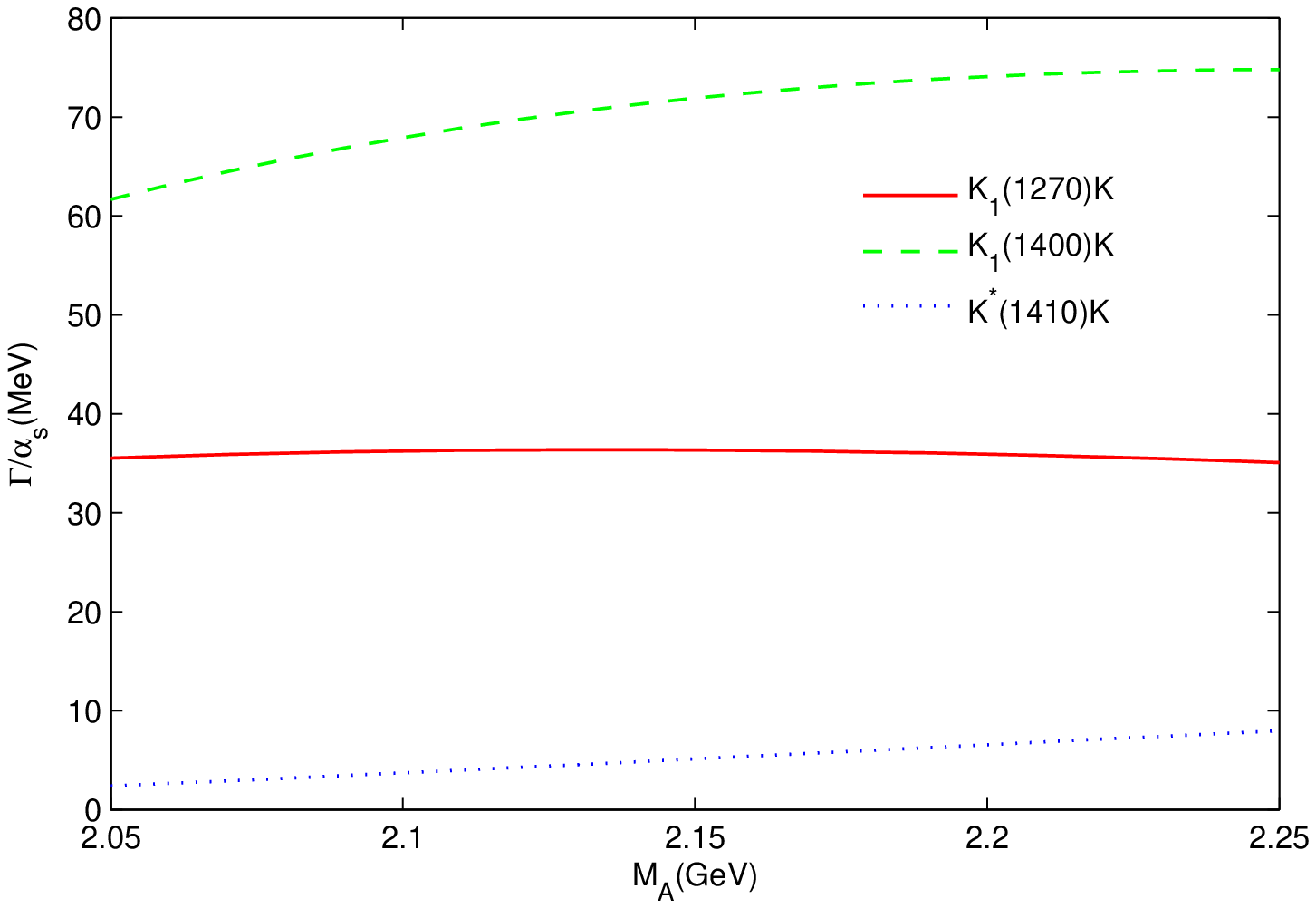}
\caption{The partial width for the $1^{--}$ gluon excited
strangeonium hybrid decaying to
$K_1(1270)K,\;K_1(1400)K,\;K^{*}(1410)K$ at various hybrid mass.}
\label{fig6}
\end{minipage}
\end{figure}

\begin{figure}[hptb]
\centering
\begin{minipage}[t]{0.47\textwidth}
\centering
\includegraphics[width=8cm]{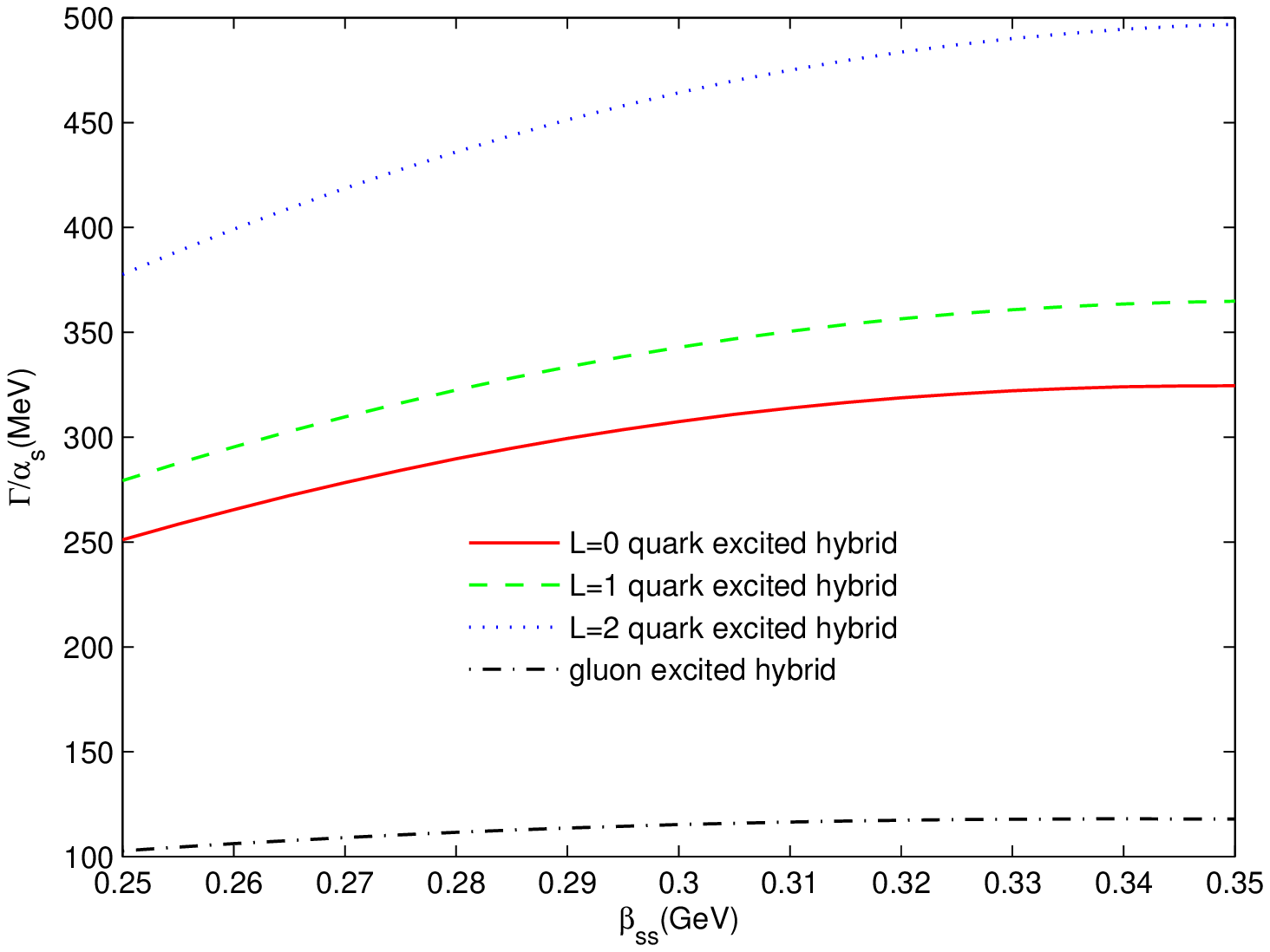}
\caption{The $\beta_{s\bar{s}}$ dependence of the total decay width
for the L=0,1,2 quark excited hybrid and the gluon excited hybrid.}
\label{fig7}
\end{minipage}%
\hspace{0.03\textwidth}%
\begin{minipage}[t]{0.47\textwidth}
\centering
\includegraphics[width=8cm]{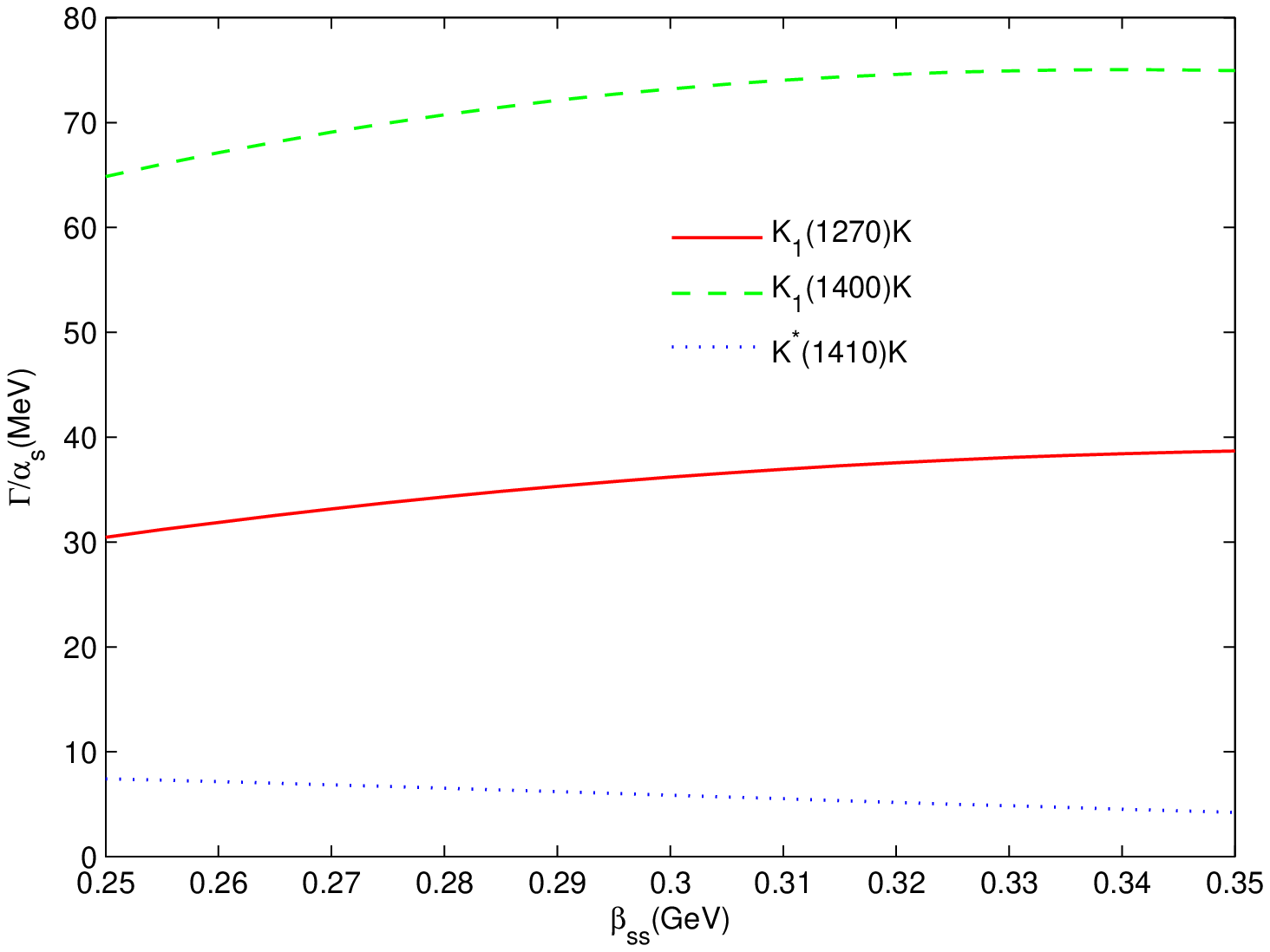}
\caption{The partial width of the $1^{--}$ gluon excited
strangeonium hybrid decaying to
$K_1(1270)K,\;K_1(1400)K,\;K^{*}(1410)K$ at various
$\beta_{s\bar{s}}$.} \label{fig8}
\end{minipage}
\end{figure}

In the case of the "gluon excited" hybrid, the allowed decay
channels and the partial decay width are shown in Table III. It
mainly decays into $K_1(1270)K$, $K_1(1400)K$ and $K^{*}(1410)K$.
Due to the selection rule, it can not decay to two pseudoscalars.
The decay width is around 122.7MeV which is consistent with the flux
tube model's prediction 148.7MeV within the uncertainties of the
models. Compared with the results in the flux tube model, the
$1^{--}$ strangeonium hybrid can not decay to $^1P_1$ state and
$\frac{\Gamma((s\bar{s}g)\rightarrow
K(1400)K)}{\Gamma((s\bar{s}g)\rightarrow K(1270)K)}\approx2$ in both
models. The decay width $\Gamma((s\bar{s}g)\rightarrow
K^{*}_2(1430)K)$ is about 0 in the constituent gluon model,
$\Gamma((s\bar{s}g)\rightarrow K^{*}_2(1430)K)\approx15$MeV in the
flux tube model, which is smaller than
$\Gamma((s\bar{s}g)\rightarrow K_1(1270)K)$ and
$\Gamma((s\bar{s}g)\rightarrow K_1(1400)K)$. So the decay patter of
the gluon excited hybrid is very similar to that in the flux tube
model, and we conclude that $Y$(2175) could mainly be the "gluon
excited" hybrid from the view of the constituent gluon model,  The
strong dependence of the decay width on the hybrid mass is displayed
in Fig.\ref{fig5}, and we show the partial width for the $1^{--}$
gluon excited strangeonium hybrid decaying to
$K_1(1270)K,\;K_1(1400)K$, $K^{*}(1410)K$ as a function of the
hybrid mass in Fig.\ref{fig6}. The $\beta_{s\bar{s}}$ dependence of
total decay width and partial decay width are shown respectively in
Fig.\ref{fig7} and Fig.\ref{fig8} in the constituent gluon model.

\begin{figure}
\centering \subfigure[]{
\label{fig:subfig:a} %% label for first subfigure
\includegraphics[width=0.17\textwidth]{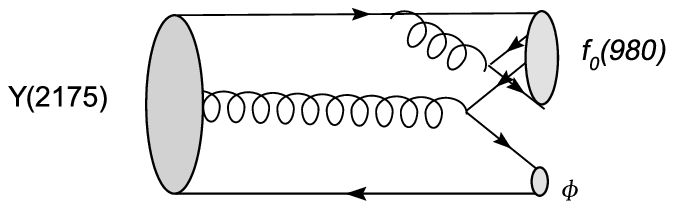}}%
\hspace{0.05in}
\subfigure[]{
\label{fig:subfig:b} %% label for second subfigure
\includegraphics[width=0.17\textwidth]{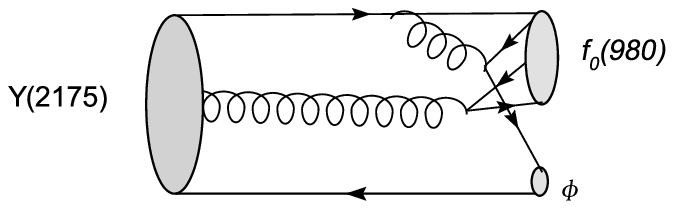}}
\centering \subfigure[]{
\label{fig:subfig:c} %% label for third subfigure
\includegraphics[width=0.17\textwidth]{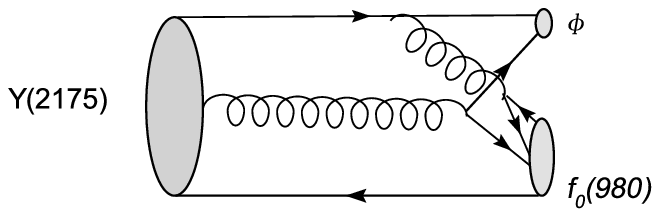}}%
\hspace{0.05in}
\centering \subfigure[]{
\label{fig:subfig:d} %% label for fourth subfigure
\includegraphics[width=0.17\textwidth]{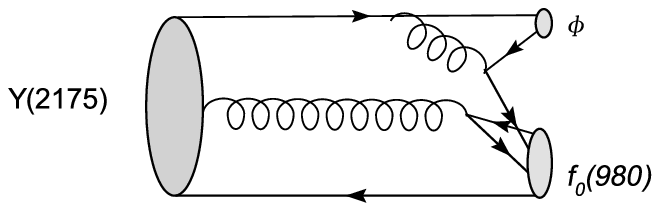}}%
\hspace{0.05in}
\centering \subfigure[]{
\label{fig:subfig:e} %% label for fifth subfigure
\includegraphics[width=0.17\textwidth]{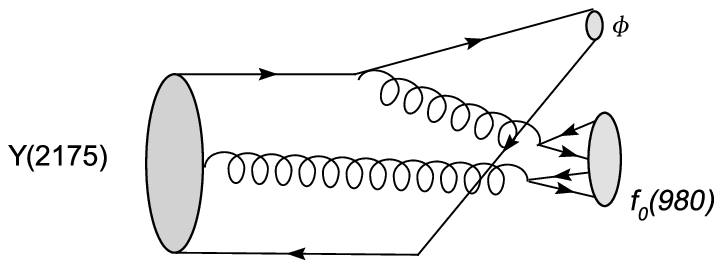}}
\caption{Decay diagrams for $Y(2175)\rightarrow\phi f_0(980)$ under the assumption that $Y$(2175)
is a $1^{--}$ strangeonium hybrid and $f_0$(980) is a four quark state.}
\label{fig9} %% label for entire figure
\end{figure}

If $Y$(2175) is a $1^{--}$ strangeonium hybrid, in the constituent
gluon model, the mechanism which generates the decay
$Y(2175)\rightarrow\phi f_0(980)\rightarrow\phi\pi\pi$ could be the
following: a gluon is emitted from the strange quark or the
anti-strange quark, both the created gluon and the constituent gluon
dissociate into a pair of quark-antiquark, then all the quarks and
antiquarks combine to form $\phi$ and $f_0(980)$. The corresponding
diagrams are shown in Fig.\ref{fig9}, where only the diagrams in
which gluon is emitted by the strange quark are plotted, and
$f_0(980)$ is assumed as a four-quark state.

\section{conclusion and discussion}

We propose that the recently observed structure $Y$(2175) is a
strangeonium hybrid, the reasonability of this hypothesis has been
studied from both the flux tube model and the constituent gluon
model. The estimate for the mass of $1^{--}$ hybrid state
$s\bar{s}g$ in both the flux tube model and the constituent gluon
model are consistent with the experimental data. In the constituent
gluon model, there exist the "quark excited" hybrids($L=0,1,2$;
$L_{s\bar{s}}=S_{s\bar{s}}=1$; $L_g$=0) and the "gluon excited"
one($L=1$; $L_g$=0; $L_{s\bar{s}}=S_{s\bar{s}}=0$) for the $1^{--}$
hybrid state. The decay width of the quark excited hybrid is very
large, it is around 250-500MeV, and the decay pattern and decay
width of the gluon excited hybrid is similar to the result of the
flux tube model. Therefore we argue that $Y$(2175) could be mainly a
"gluon excited" strangeonium hybrid from the view of constituent
gluon. Furthermore, both models predict the decay width is about
100-150MeV with some theoretical uncertainties, which is consistent
with the present data within errors.

Both the flux tube model and the constituent gluon model(for the
"gluon excited" hybrid) predict that the $1^{--}$ hybrid $s\bar{s}g$
dominantly decays into $K_1(1400)K$ and $K_1(1270)K$, which is
consistent with the well-known selection rule of hybrid decay. So
the experimental search of channels $Y(2175)\rightarrow
K_1(1400)K\rightarrow\pi K^{*}(892)K$, $Y(2175)\rightarrow
K_1(1270)K\rightarrow\rho KK$ and $Y(2175)\rightarrow
K_1(1270)K\rightarrow\pi K_0^*(1430)K$ are suggested, which is a
important test of our scenario. However, $Y$(2175) dominantly decays
into $\phi\eta$ and $\phi\eta'$ in the tetraquark picture, the
search for $Y(2175)\rightarrow\phi\eta$ and
$Y(2175)\rightarrow\phi\eta'$ is also merited to discriminate the
two interpretations, although the tetraquark hypothesis is not
favored by the available experimental information.

Although both $Y$(2175) and $Y$(4260) have been found in the ISR
$e^{+}e^{-}$ annihilation by the Babar collaboration, their decay
property is different. In the hybrid picture, they both prefer to
decay into the final states with a P-wave meson because of the "S+P"
selection rule, i.e., $K_1(1400)K$ and $K_1(1270)K$ are the dominant
decay modes as we have shown above, and $Y$(4260) has strong
coupling to $D\overline{D}^{\;**}$ and $\overline{D}D^{**}$.
However, $Y$(4260) are slightly larger than the threshold of
$D\overline{D}^{\;**}$ and $\overline{D}D^{**}$, so those decay
modes are suppressed. The $K^{+}K^{-}\pi^{+}\pi^{-}$ final state in
the ISR $e^{+}e^{-}$ annihilation has been analyzed by the Babar
collaboration\cite{fbabar}, it is essential to note that a broad
structure in the region of the $K_1(1270)$ and $K_1(1400)$ is
claimed by performing the three-body mass combination, which is
shown in Fig. 19(c) in that paper, this signal seems to support our
hybrid picture of Y(2175). It is essential to perform the mass
combination of $K_1(1270)K$, $K^{*}_2 (1430)K$ and $K_1(1400)K$,
perform the partial wave analysis, and measure the branching ratios
etc. in order to verify the hybrid hypothesis of $Y$(2175).

No matter $Y$(2175) is a hybrid or tetraquark quark state, it can
mix with other conventional mesons, which have the same quantum
numbers as $Y$(2175). However, the nearest $1^{--}$ isospin singlet
to $Y$(2175) is $\phi(1680)$\cite{pdg}, and quantum mechanics tells
us that the mixing strength is related to the inverse of the energy
difference of the two states, so the mixing effect is expected to be
rather small.

To confirm $Y$(2175) being a hybrid, further deep theoretical
understanding and predictions about the properties of $1^{--}$$
(s\bar{s}g)$ hybrid state are needed to be confronted with the
experimental data. It is essential to investigate if $Y$(2175) could
be a conventional strange quarkonium, {\it e.g.} $2{\;^3D_1}$ or
$3{\;^3S_1}$ $s\bar{s}$ quarkonium. The decay of $3{\;^3S_1}$
$s\bar{s}$ quarkonium has been studied in $^3P_0$ model by T.Barnes
{\it et al.,}\cite{barnes1}, and this state is predicted to be a
rather broad state , $\Gamma\approx 380$ MeV, so $Y(2175)$ should
not be $3{\;^3S_1}$ $s\bar{s}$ state. However, the decay of
$2{\;^3D_1}$ $s\bar{s}$ quarkonium has not been considered so far,
it is urgent and interesting to study $2{\;^3D_1}$ $s\bar{s}$ state
from quark model\cite{ding}. Experimentally, confirmation and
dictated studies of $Y$(2175) at BES and CLEO is valuable.

%\begin{figure}[hptb]
%\begin{center}
%\includegraphics[width=0.80\textwidth]{VA.eps}
%\caption{The decay width of the $1^{--}$ strangeonium hybrid at
%various hybrid mass}
%\end{center}
%\end{figure}
%\begin{figure}[hptb]
%\begin{center}
%\includegraphics[width=0.80\textwidth]{PK.eps}
%\caption{The partial width of the $1^{--}$ gluon excited
%strangeonium hybrid to $K_1(1270)K,K_1(1400)K$ at various hybrid
%mass}
%\end{center}
%\end{figure}

\section *{ACKNOWLEDGEMENTS}
\indent We acknowledge Professor T.Barnes for his professional
suggestion of studying if $Y(2175)$ could be a $2{\;^3D_1}$
$s\bar{s}$ quarkonium. We are indebted to Dr. Jie-Jie Zhu for
helpful discussion. This work is partially supported by National
Natural Science Foundation of China under Grant Numbers 90403021,
and by the PhD Program Funds 20020358040 of the Education Ministry
of China and KJCX2-SW-N10 of the Chinese Academy.

\begin{appendix}
\section{the analytical expressions for $h_0,g_1,h_2$ }
$h_0,g_1,h_2$ are as follows:
\begin{equation}
\label{a1}h_0=\tilde{\beta}^2\int_0^{\infty}dr\;r^{3+\delta}j_{0}(\frac{Mp_Br}{m+M})\exp(-(2\beta^2_A+\tilde{\beta}^2)\frac{r^2}{4})
\end{equation}
\begin{equation}
\label{a2}g_1=\frac{2mp_B}{m+M}\int^{\infty}_0dr\;r^{2+\delta}j_{1}(\frac{Mp_Br}{m+M})\exp(-(2\beta^2_A+\tilde{\beta}^2)\frac{r^2}{4})
\end{equation}
\begin{equation}
\label{a3}h_2=\tilde{\beta}^2\int_0^{\infty}dr\;r^{3+\delta}j_{2}(\frac{Mp_Br}{m+M})\exp(-(2\beta^2_A+\tilde{\beta}^2)\frac{r^2}{4})
\end{equation}
In the above formula, $M$ and $m$ are respectively the masses of the
original quark and the created quark. The integrals can be evaluated
in terms of the confluent hypergeometric functions using the general
formula
\begin{equation}
\label{a4}\int_0^{\infty}dr\;r^nj_{m}(ar)e^{-br^2}=\sqrt{\pi}\;\frac{a^m}{b^{\phi}}\frac{\Gamma[\phi]}{2^{m+2}\Gamma[m+3/2]}\;{_1F_1}(\phi,m+\frac{3}{2},-\frac{a^2}{4b})
\end{equation}
where $\phi=(m+n+1)/2$, then
\begin{equation}
\label{a5}h_0=\frac{2^{3+\delta}\tilde{\beta}^2}{(2\beta^2_A+\tilde{\beta}^2)^{2+\delta/2}}\Gamma(2+\frac{\delta}{2})\;{_1F_1}(2+\frac{\delta}{2},\frac{3}{2},-\frac{M^2p^2_B}{(m+M)^2(2\beta^2_A+\tilde{\beta}^2)})
\end{equation}
\begin{equation}
\label{a6}g_1=\frac{mMp^2_B}{(m+M)^2}\frac{2^{4+\delta}}{3(2\beta^2_A+\tilde{\beta}^2)^{2+\delta/2}}\Gamma(2+\frac{\delta}{2})\;{_1F_1}(2+\frac{\delta}{2},\frac{5}{2},-\frac{M^2p^2_B}{(m+M)^2(2\beta^2_A+\tilde{\beta}^2)})
\end{equation}
\begin{equation}
\label{a7}h_2=\frac{(Mp_B)^2}{(m+M)^2}\frac{2^{5+\delta}\tilde{\beta}^2}{15(2\beta^2_A+\tilde{\beta}^2)^{3+\delta/2}}\Gamma(3+\frac{\delta}{2})\;{_1F_1}(3+\frac{\delta}{2},\frac{7}{2},-\frac{M^2p^2_B}{(m+M)^2(2\beta^2_A+\tilde{\beta}^2)})
\end{equation}

\section{the partial decay width $M_{\ell,J}$ in the constituent gluon model}
Starting from the interaction Hamiltonian Eq.(\ref{10}) and the
non-relativistic wavefunction Eq.(\ref{12}) and Eq.(\ref{13}), the
matrix element $\langle
BC|H_I|A\rangle=g_s(2\pi)^3\delta^{3}(\mathbf{p}_A-\mathbf{p}_B-\mathbf{p}_C)\;M_{\ell,J}(A\rightarrow
BC)$ can be easily obtained, and the partial wave amplitude
$M_{\ell,J}$ has the following form:
\begin{eqnarray}
\nonumber&& M_{\ell J}(A\rightarrow BC)=\sum_{\begin{array}{cccccc}
M_{L_g},&\lambda_g,&M_{L_{s\bar{s}}},&M_{S_{s\bar{s}}},&M_{L_B},&M_{S_{B}},\\
M_{L_C},&M_{S_C},&M_{J_B},&M_{J_C},&M_{J},&M_{\ell}\\
\end{array}}{\cal C}\;{\cal F}\;
{\cal S}(M_{S_{s\bar{s}}},\lambda_g,M_{S_B},M_{S_{C}})\\
\nonumber&&I(M_{L_{s\bar{s}}},M_{L_{g}},M_{L_B},M_{L_{C}},M_{\ell})\langle
L_g,M_{L_g};1,\lambda_{g}|J_{g},M_{L_{g}}+\lambda_g\rangle\langle
L_{s\bar{s}},M_{L_{s\bar{s}}};J_{g},M_{L_{g}}+\lambda_g|L,M_{L_{g}}\\
\nonumber&&+\lambda_g+M_{L_{s\bar{s}}}\rangle\langle
L,M_{L_{g}}+\lambda_g+M_{L_{s\bar{s}}}; S_{s\bar{s}},
M_{S_{s\bar{s}}}|J_A, M_{J_A}\rangle\langle
L_B,M_{L_B};S_{B},M_{S_B}|J_B,M_{J_B}\rangle\langle
L_C,M_{L_C};\\
\label{14}&&S_C,M_{S_C}|J_C,M_{J_C}\rangle\langle
J_B,M_{J_B};J_{C},M_{J_C}|J,M_J\rangle\langle\ell,M_{\ell};J,M_J|J_A,M_{J_A}\rangle
\end{eqnarray}
Here ${\cal C},\;{\cal F},\;{\cal
S}(M_{S_{s\bar{s}}},\lambda_g,M_{S_B},M_{S_{C}})$ and
$I(M_{L_{s\bar{s}}},M_{L_{g}},M_{L_B},M_{L_{C}},M_{\ell})$ are
respectively color, flavor, spin and spatial overlap factors with
${\cal C}=\frac{2}{3}$. In the non-relativistic limit, the spin
overlap factor is:
\begin{eqnarray}
\nonumber&&{\cal
S}(M_{S_{s\bar{s}}},\lambda_g,M_{S_B},M_{S_C})=\sum_{S}\sqrt{6(2S_B+1)(2S_C+1)(2S_{s\bar{s}}+1)}\left
\{\begin{array}{lll}
\frac{1}{2}&\frac{1}{2} & S_B\\
\frac{1}{2}&\frac{1}{2} & S_C\\
S_{s\bar{s}}& 1 & S
\end{array}
\right\}\langle
S_{s\bar{s}},M_{S_{s\bar{s}}};1,\\
\label{15}&&\lambda_g|S,M_{S_{B}}+M_{S_C}\rangle\langle
S_B,M_{S_B};S_C,M_{S_C}|S,M_{S_{B}}+M_{S_C}\rangle
\end{eqnarray}
The flavor overlap factor ${\cal F}$ is:
\begin{eqnarray}
\label{16}&&{\cal F}=\sqrt{(2I_B+1)(2I_C+1)(2I_{A}+1)}\left
\{\begin{array}{lll}
i_1& i_3 & I_B\\
i_2&i_{4} & I_C\\
I_A& 0 & I_A
\end{array}
\right\}\eta\;\varepsilon
\end{eqnarray}
where $\eta=1$ if the gluon goes into strange quarks and
$\eta=\sqrt{2}$ if it goes into non-strange ones. $\varepsilon$ is
the number of the diagrams contributing to the decay. Finally the
spatial overlap is given by:
\begin{eqnarray}
\nonumber&&I(M_{L_{s\bar{s}}},M_{L_{g}},M_{L_B},M_{L_{C}},M_{\ell})=\int\int\int\frac{d^3\mathbf{p}d^3\mathbf{k}}{\sqrt{2\omega}(2\pi)^6}\;\psi_{L_{s\bar{s}} M_{L_{s\bar{s}}}}(\mathbf{p}_B-\mathbf{p})\;\psi_{L_gM_{L_g}}(\mathbf{k})\;\psi^*_{L_B M_{L_B}}(\frac{m\mathbf{p}_B}{M+m}\\
\label{17}&&-\mathbf{p}-\frac{\mathbf{k}}{2})\;\psi^{*}_{L_C
M_{L_C}}(-\frac{m\mathbf{p}_B}{M+m}+\mathbf{p}-\frac{\mathbf{k}}{2})\;d\Omega_{B}{Y^{M_{\ell}}_{\ell}}^{*}(\Omega_{B})
\end{eqnarray}
The partial decay width is:
\begin{equation}
\label{18}\Gamma_{\ell J}(A\rightarrow
BC)=\frac{\alpha_s}{\pi}\frac{p_BE_BE_C}{M_A}|M_{\ell
J}(A\rightarrow BC)|^2
\end{equation}
where $M_{\ell J}(A\rightarrow BC)$ is the partial wave amplitude,
and $\Gamma_{\ell J}$ is the partial width of that partial
wave\cite{partialwave}. As usual, we would like to use the S.H.O
basis wavefunctions, thereby enabling analytic studies that reveal
the relationship among amplitudes.
\begin{eqnarray}
\label{19}\psi_{L_{s\bar{s}}
M_{L_{s\bar{s}}}}(\mathbf{p})=[\frac{16\pi^3}{\Gamma(\frac{3}{2}+L_{s\bar{s}})\;\beta_{s\bar{s}}^{2L_{s\bar{s}}+3}}]^{1/2}\;{\cal
Y}^{M_{L_{s\bar{s}}}}_{L_{s\bar{s}}}(\mathbf{p})\exp[-\frac{\mathbf{p}^2}{2\beta^2_{s\bar{s}}}]
\end{eqnarray}
and
\begin{equation}
\label{20}\psi_{L_g
M_{L_g}}(\mathbf{p}_g)=[\frac{16\pi^3}{\Gamma(\frac{3}{2}+L_{g})\;\beta_g^{2L_{g}+3}}]^{1/2}\;{\cal
Y}^{M_{L_g}}_{L_g}(\mathbf{p}_g)\exp[-\frac{\mathbf{p}^2_g}{2\beta^2_g}]
\end{equation}
here ${\cal
Y}^{M_{L_{s\bar{s}}}}_{L_{s\bar{s}}}(\mathbf{p})=|\mathbf{p}|^{L_{s\bar{s}}}Y^{M_{L_{s\bar{s}}}}_{L_{s\bar{s}}}(\mathbf{\hat{p}})$
is a solid harmonic polynomial, and analogously for ${\cal
Y}^{M_{L_g}}_{L_g}(\mathbf{p}_g)$. The wavefunctions $\psi_{L_B
M_{L_B}}$ and $\psi_{L_C M_{L_C}}$ are defined similarly. In the
following, we will take $\beta_B=\beta_C$. If the final states are
two $S-$wave mesons, ${\it i.e}, L_B=L_C=0$, then the overlap
integral Eq.(\ref{17}) can be integrated out analytically.
\begin{eqnarray}
\nonumber&&I(M_{L_{s\bar{s}}},M_{L_{g}},0,0,M_{\ell})=\frac{2}{\sqrt{\pi\omega}\;\beta^3_B}[\frac{1}{\Gamma(L_{s\bar{s}}+\frac{3}{2})\Gamma(L_g+\frac{3}{2})\beta^{2L_{s\bar{s}}+3}_{s\bar{s}}\beta^{2L_g+3}_g}]^{1/2}[\frac{2\pi\;\beta^2_{s\bar{s}}\beta^{2}_{g}\beta^4_{B}}{(\beta^2_{B}+2\beta^2_{s\bar{s}})(\beta^2_B+\beta^2_g/2)}]^{3/2}\\
\label{21}&&[\frac{2\;\beta^2_{s\bar{s}}p_B}{\beta^2_{B}+2\beta^2_{s\bar{s}}}\frac{M}{M+m}]^{L_{s\bar{s}}}\exp[-\frac{M^2}{(M+m)^2}\frac{p^2_{B}}{\beta^2_{B}+2\beta^2_{s\bar{s}}}]\;\delta_{\ell,L_{s\bar{s}}}\;\delta_{M_{\ell},M_{L_{s\bar{s}}}}\delta_{L_g,0}\;\delta_{M_g,0}
\end{eqnarray}
The above equation implies that the "gluon excited" hybrid doesn't
decay into two ground states mesons, which has been shown to be true
in both the flux tube model and the constituent gluon
model\cite{isgur,orsay,page2}. Particularly for the quark excited
hybrid $L_g=0,L_{s\bar{s}}=1$,
\begin{eqnarray}
\nonumber&&I(M_{L_{s\bar{s}}},0,0,0,M_{\ell})=\sqrt{\frac{\pi}{3\omega}}\frac{\beta^{5/2}_{s\bar{s}}\beta^{3/2}_g\beta^3_{B}}{(\beta^2_B/2+\beta^2_g/4)^{3/2}(\beta^2_{B}/2+\beta^2_{s\bar{s}})^{5/2}}\frac{2Mp_B}{M+m}\exp[-\frac{M^2}{(M+m)^2}\frac{p^2_B}{\beta^2_{B}+2\beta^2_{s\bar{s}}}]\\
\label{22}&&\times\delta_{\ell,L_{s\bar{s}}}\;\delta_{M,M_{L_{s\bar{s}}}}
\end{eqnarray}
If the final states are $S+P-$wave meson pairs, the overlap integral
Eq.(\ref{17}) can be exactly integrated out likewise. For the gluon
excited hybrid $L_{s\bar{s}}=0,L_g=1$, the relevant overlap integral
is:
\begin{eqnarray}
\nonumber&&I(0,M_{L_g},M_{L_B},0,M_{\ell})=-\sqrt{\frac{\pi}{2\omega}}\frac{\beta^{3/2}_{s\bar{s}}\beta^{5/2}_{g}\beta^4_B}{(\beta^2_{B}/2+\beta^2_{s\bar{s}})^{3/2}(\beta^2_B/2+\beta^2_g/4)^{5/2}}\exp[-\frac{M^2}{(M+m)^2}\frac{p^2_B}{\beta^2_{B}+2\beta^2_{s\bar{s}}}]\\
\label{23}&&\times\delta_{M_{L_g},M_{L_B}}\delta_{\ell,0}\;\delta_{M_{\ell},0}
\end{eqnarray}

where the $\delta_{\ell,0}$ term above tells us that the $1^{--}$
hybrid decay into $S+P-$wave final states in relative $S-$wave, so
the $1^{--}$ strangeonium hybrid can not decay to $K^{*}_2(1430)K$
due to conservation of angular momentum. And the corresponding
overlap integral for the quark excited hybrid is:
\begin{eqnarray}
\nonumber&&I(M_{L_{s\bar{s}}},0,M_{L_B},0,M_{\ell})=\sqrt{\frac{2\pi}{\omega}}\frac{\beta^{5/2}_{s\bar{s}}\beta^{3/2}_{g}\beta^4_{B}}{(\beta^2_{B}/2+\beta^2_{s\bar{s}})^{5/2}\;(\beta^2_B/2+\beta^2_g/4)^{3/2}}\exp[-\frac{M^2}{(M+m)^2}\frac{p^2_B}{\beta^2_{B}+2\beta^2_{s\bar{s}}}]\\
\nonumber&&\times[\delta_{M_{L_{s\bar{s}}},M_{L_B}}\delta_{\ell,0}\;\delta_{M_{\ell},0}-\frac{2}{3}\sqrt{2\ell+1}\;(\frac{M}{M+m})^2\frac{p^2_B}{\beta^2_{B}+2\beta^2_{s\bar{s}}}\langle1,M_{L_B};\ell,
M_{\ell}|1,M_{L_{s\bar{s}}}\rangle\langle1,0;\ell,0|1,0\rangle]
\end{eqnarray}

Both the quark excited hybrid and the gluon excited hybrid can decay
into $2S+1S$ final states, if these processes are not forbidden by
the phase space. In order to distinguish the notations from those
for hybrid decaying into $S+S$ and $S+P$ final states, the overlap
integral for hybrid decaying into $2S+1S$ final states is denoted as
$I^{\prime}(M_{L_{s\bar{s}}},M_{L_g},0,0,M)$, which is of the
following form,
\begin{eqnarray}
\nonumber&&I^{\prime}(M_{L_{s\bar{s}}},0,0,0,M_{\ell})=-\sqrt{\frac{\pi}{\omega}}\;\frac{\beta^{5/2}_{s\bar{s}}\beta^{3/2}_g\beta^{3}_B}{(\beta^2_B+\beta^2_g/2)^{5/2}\;(\beta^2_{B}/2+\beta^2_{s\bar{s}})^{9/2}}\;\frac{Mp_B}{3(M+m)^3}\exp[-\frac{M^2}{(M+m)^2}\\
\nonumber&&\frac{p^2_{B}}{\beta^2_{B}+2\beta^2_{s\bar{s}}}]\{-m^2(\beta^2_B+2\beta^2_{s\bar{s}})(5\beta^4_{B}+\beta^2_{B}\beta^2_{g}-3\beta^2_{g}\beta^2_{s\bar{s}})-2mM(\beta^2_{B}+2\beta^2_{s\bar{s}})(5\beta^4_B+\beta^2_B\beta^2_{g}-3\beta^2_{g}\beta^2_{s\bar{s}})\\
\label{add3}&&+M^2[\;p^2_B\beta^2_B(2\beta^2_B+\beta^2_g)-(\beta^2_B+2\beta^2_{s\bar{s}})(5\beta^4_B+\beta^2_B\beta^2_g-3\beta^2_g\beta^2_{s\bar{s}})]\}\delta_{L_g,0}\;\delta_{\ell,L_{s\bar{s}}}\;\delta_{M_{\ell},M_{L_{s\bar{s}}}}
\end{eqnarray}
and:
\begin{eqnarray}
\nonumber&&I^{\prime}(0,M_{L_{g}},0,0,M_{\ell})=-\sqrt{\frac{\pi}{\omega}}\;\frac{\beta^{3/2}_{s\bar{s}}\beta^{5/2}_g\beta^{5}_B}{(\beta^2_B+\beta^2_g/2)^{5/2}(\beta^2_{B}/2+\beta^2_{s\bar{s}})^{5/2}}\;\frac{4Mp_B}{3(M+m)}\exp[-\frac{M^2}{(M+m)^2}\\
\label{add4}&&\frac{p^2_{B}}{\beta^2_{B}+2\beta^2_{s\bar{s}}}]\;\delta_{L_{s\bar{s}},0}\;\delta_{\ell,L_{g}}\;\delta_{M_{\ell},M_{L_{g}}}
\end{eqnarray}
The first result Eq.(\ref{add3}) corresponds to the "quark excited"
hybrid, and the second Eq.(\ref{add4}) corresponds to the "gluon
excited" hybrid.

\end{appendix}

%\begin{figure}[hptb]
%\begin{center}
%\includegraphics*[5pt,580pt][400pt,800pt]{f3.eps}
%\caption{The configuration in common quark model, where gray circles
%denote quarks and red circles denote antiquarks}.
%\end{center}
%\end{figure}

%\begin{figure}[hptb]
%\begin{center}
%\caption{The partial width for the strangeonium hybrid decay as a
%function of the hybrid mass.}
%\includegraphics*[width=8cm]{ALLCHAN.eps}
%\includegraphics*[width=8cm]{WIDTH.eps}
%\end{center}
%\end{figure}

%\begin{figure}
%\centering \subfigure[ The partial width for the strangeonium hybrid
%decay as a function of the hybrid mass]{
%\label{fig:subfig:a} %% label for first subfigure
%\includegraphics[width=0.45\textwidth]{ALLCHAN.eps}}%
%\hspace{1in}
%\subfigure[ The total width of the strangeonium hybrid as a function
%of the hybrid mass]{
%\label{fig:subfig:b} %% label for second subfigure
%\includegraphics[width=0.45\textwidth]{WIDTH.eps}}
%\caption{Two Subfigures}
%\label{fig:subfig} %% label for entire figure
%\end{figure}

\end{document}